# Mobile Multimedia Recommendation in Smart Communities: A Survey

Feng Xia, *Senior Member, IEEE*, Nana Yaw Asabere, Ahmedin Mohammed Ahmed, Jing Li, and Xiangjie Kong, *Member, IEEE*

*Abstract*—Due to the rapid growth of internet broadband access and proliferation of modern mobile devices, various types of multimedia (e.g. text, images, audios and videos) have become ubiquitously available anytime. Mobile device users usually store and use multimedia contents based on their personal interests and preferences. Mobile device challenges such as storage limitation have however introduced the problem of mobile multimedia overload to users. In order to tackle this problem, researchers have developed various techniques that recommend multimedia for mobile users. In this survey paper, we examine the importance of mobile multimedia recommendation systems from the perspective of three smart communities, namely, mobile social learning, mobile event guide and context-aware services. A cautious analysis of existing research reveals that the implementation of proactive, sensor-based and hybrid recommender systems can improve mobile multimedia recommendations. Nevertheless, there are still challenges and open issues such as the incorporation of context and social properties, which need to be tackled in order to generate accurate and trustworthy mobile multimedia recommendations.

*Index Terms*—Context-awareness, mobile event guide, mobile social learning, multimedia recommender systems, smart community

## I. Introduction

THE current growth and tremendous proliferation of mobile devices has paved the way for accessing and capturing different types of multimedia contents (e.g. text, images, audios and videos). Generations of mobile devices started from analogue and migrated to digital. Mobile multimedia contents began with text and later generated to still images, animation, audio and video. Currently, different types of multimedia contents are available due to diverse generations of mobile devices, namely, the first generation (1G), second generation (2G), third generation (3G) and forth generation (4G) [1]-[3].

As a result of technological trends and changes, modern computers are becoming portable, smaller and thus mobile. These modern computers are equipped with higher performance in terms of computational processing speed, memory size etc. Some examples of new forms of computers include smartphones, tablets, and personal digital assistants (PDAs). Many physical things will possess computing and communication capabilities of different levels, which are provided by small and perhaps invisible computers embedded therein. This resultant incorporation of networked computing and physical dynamics has led to the emergence of Cyber-Physical Systems (CPS), which have become a very interesting area for research in recent years.

The growing trends towards the convergence of CPS and social computing have led to the emergence of smart communities. They are usually composed of physical objects/things and human beings that interact and cooperate with each other. Smart communities promise to support a number of state-of-the-art applications and services that will improve the quality of life [4][5].

Mobile devices are well on their way of becoming personal and sensing platforms. They were primarily used solely as communication devices but can now be used to access and capture multimedia contents for wide range of activities in smart communities [2][6]. In academia and education, mobile devices are used for mobile social learning. During events such as conferences and workshops, mobile devices can be used as an event guides for the planner, organizers and attendees. In other activities which involve context-aware services such as entertainment, shopping, tourism and location-based activities, mobile devices play different and important roles for individual users. Mobile devices used in such smart communities have limited storage and can therefore not store a lot of multimedia contents for users. Nowadays multimedia contents are becoming more prevalent in mobile devices, making them an advantageous gateway for the individual.

In order to curb multimedia information overload and to allow users to have access to relevant multimedia contents in their mobile devices, today's main focus and challenges of researchers is on how to develop multimedia recommender systems for mobile devices. Furthermore, users of mobile devices in smart communities have different interest, preferences, tastes and demography and would usually like to store multimedia contents that are only relevant to them. It is however not easy for users to search or manage large volumes of multimedia contents in a mobile device with limited resources such as storage. To corroborate the importance of

This work was partially supported by the Natural Science Foundation of China under Grant No. 60903153, No. 61203165, and No. 61174174, Liaoning Provincial Natural Science Foundation of China under Grant No. 201202032, and the Fundamental Research Funds for the Central Universities.
The authors are with School of Software, Dalian University of Technology, Dalian 116620, China.
Corresponding author: Xiangjie Kong; E-mail: xjkong@ieee.org

recommender systems in smart communities so that the aforementioned problems and challenges in mobile multimedia can be solved, some researchers [1][7]-[20] have investigated and developed mobile multimedia recommender systems in different smart communities.

A mobile multimedia recommendation system solves the problem of multimedia overload in mobile devices by predicting and presenting relevant multimedia information to users. Users therefore don't waste appropriate time searching for contents that they are/might be interested in. The increasing interest of multimedia contents in different smart communities and the growing usage of mobile devices motivates us to conduct a detailed analysis and bring to light some important existing research work and relevant open research issues in this important scientific area.

In this paper, we survey the importance and relevance of mobile multimedia recommendation systems for three smart communities, namely, mobile social learning, mobile event guides and context-aware services. Furthermore, we present an overview of mobile multimedia recommendation systems from the perspective of applications, architectures and algorithms. Through careful examination of the state-of-the-art, we outline challenges and open issues that require attention so that recommendations generated for mobile users will be more accurate, reliable, efficient and trustworthy.

The rest of this paper is organized as follows. In Section II we present an overview of the smart communities. We examine the taxonomy of mobile multimedia recommender systems in Section III. Section IV reviews the existing research of mobile multimedia recommender systems for smart communities. Discussions and open issues are presented in Section V and then Section VI concludes the paper.

## II. SMART COMMUNITIES

A smart community can be roughly understood as a group of connected social objects that interact with each other over ubiquitous networks and deliver smart services to all members [4]. Smart communities continuously monitor the social community environment from various aspects; and when necessary, automatic or human-controlled physical feedback is input to improve social community interests, safety and emergency response abilities.

The members of a smart community are objects that can be human individuals, as well as physical things such as a desk, a watch, a pencil, a door and a key. It is also possible that some other living things (i.e. besides human beings) might be included, for example, a plant and a cat. In most cases, these objects have implicit links among them. As a result of factors such as societal challenges of the elderly and extraordinary population ageing with the unavoidable consequences related to disability and care issues as well as recent technologies resulting in the combination of both CPS and social computing, building smart communities is very important for a society.

Xia and Ma [4] discussed notable technical challenges of building smart communities such as community design and management, autonomous networking, ubiquitous sensing and collaborative reasoning and modeling that exist when building smart communities. Technical challenges such as cooperative authentication and detecting unreliable nodes when building smart communities were discussed in Li *et al.* [5]. They also discussed application challenges comprising of target tracking and intrusion detection when building smart communities. Figure 1 depicts a general view of a smart community featuring integration of the cyber, physical, and social worlds.

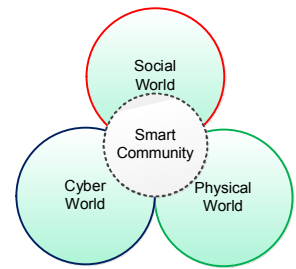

Figure 1. High-level view of a smart community

The global proliferation of wireless and mobile technologies coupled with mobile devices depicts the importance of mobile computing in smart communities. As mentioned above, this survey paper therefore pays special attention and describes the following smart communities one by one.

*A. Mobile Social Learning*

Recent developments in the field of Information and Communication Technology (ICT), specifically mobile computing technologies have led to a renewed interest among educators in using mobile devices for learning [21]-[24]. Modern mobile devices have many of the features of desktop computers with access to broadband internet networks as tools for teaching and learning.

Mobile learning is an ongoing learning landscape and education mode in which users can use mobile communication terminals to assist them to learn. Mobile learning is formed in the background of knowledge exploration. The main feature of mobile learning is mobility. It breaks through the bottleneck of traditional distance learning and e-learning and extends the scope of network multimedia distance education. Mobile learning can provide learners with the learning autonomy, and also provide the instructors and education administrators with more flexible teaching and management methods. Lifelong learning can be realized through mobile learning.

Generally, traditional teaching methods have numerous drawbacks. One of them is the fact that students attend a course, take notes and leave without any collaboration in the classroom due to e.g. lack of lecture/classroom time or the non-realization of the importance of collaborative learning. Social learning tries to solve this ineffectiveness. It is an educational method in which students work together in small groups towards a common goal [25][26]. The teacher acts as a coach, mentor or facilitator of the learning process. The successful achievement of the common goal is shared among all group members. The students take initiative and responsibility for learning. They actively learn by doing, by practice and by experience. Mobile social learning is student-centered, task-based, activity-based learning approaches that provide several advantages to the student [27].

The weaknesses associated with PC-based collaboration can be addressed by mobile social learning [28]. PC has been designed for personal use, with the expectation that learners sit behind a computer screen. Mobility allows learners to have the physical control and interactivity in their collaborative work since they can carry the mobile device while they establish a face-to-face interaction. Learners can even access internet resources or communicate with another learner or expert at a distant location. Using a mobile device, learners can regain the benefits of 'natural' face-to-face collaboration with their peers and also interact with each other.

While traditional Computer Supported Collaborative Learning (CSCL) refers to PC-based collaboration for physically dispersed learners, mobile social learning allows learners in face-to-face collaboration to enjoy computing support using mobile devices. Further advancement of mobile technologies may even open up the exciting possibilities of allowing co-located team members to interact and learn seamlessly with dispersed peers. When this happens, mobile devices can be said to be supporting the convergence of distance collaborative learning with face-to-face collaborative learning. In particular, mobile learning devices would be supporting learners to act in the physical world, access symbolic internet resources, capture learning experiences and interact with others locally and at a distance.

In a mobile social learning community, there are different learners and learning groups with different interests and preferences of programmes, courses and subjects [27]. Learners in a mobile social learning community who use multimedia as part of their learning process usually store multimedia contents they prefer and are interested. Such storages really depend on the storage capacities of their mobile devices.

Clearly, recommendation systems in mobile computing are very necessary for filtering required academic multimedia resources for learners in mobile social learning in order to meet their multimedia learning interests and preferences. Figure 2 shows a diagram of a mobile social learning smart community involving a teacher and three different groups of students/learners. The server which has a mobile Learning Management System (mLMS) is directly connected to the teachers and students/learners as clients through wireless internet connectivity.

*B. Mobile Event Guide*

Informally, an event is an information item that is only valid for a short period of time. These kinds of events (e.g., meetings, conferences, tradeshows, festivals, entertainment and so on) have been regularly organized worldwide each year. The organizing process consists of sequences of major activities involving several distant participants. The characteristics of the event management process, therefore, suit the solution of distributed systems. Managing an event for a smart community can be a frustrating process for both organizers and attendees.

However, technological advances in wireless networks and the advent of new mobile devices have made the development of mobile applications possible so that both parties can effectively be supported. Mobile applications are easily accessible through a smart phone and the user can effortlessly upload or retrieve any pertinent information. One of these applications for the smart community is mobile event guide that can help in meeting and conference planning, tradeshows and event organizing in smart communities.

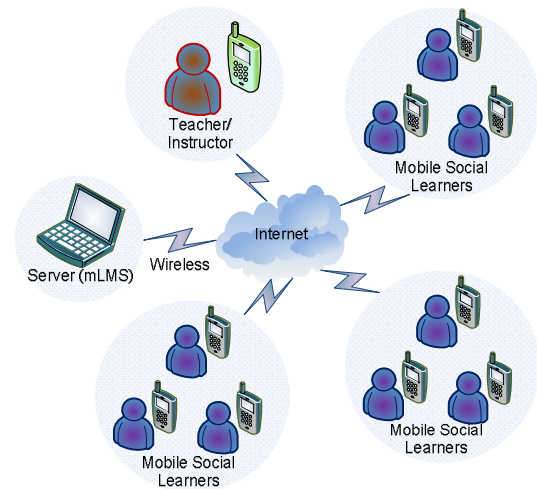

Figure 2. A mobile social learning smart community

The rapid innovation and user adoption of sophisticated smart phone technology has certainly played a role in this transition. Despite the emergence of mobile technology, however, the majority of events still use paper-based event guides and programs. As a result of the advent of smart mobile devices, that is all about to change. Many events have been regularly organized worldwide each year. The organizing process consists of sequences of major activities involving several distant participants. The characteristics of an event management process, therefore, suit the solution of distributed systems.

Mobile event guide is one of the example applications for multimedia recommendation since it includes very important services like audio and video systems development for an event and distributing or releasing it for interested participants of the event in the community. Different participants and attendees of events such as a conference are likely to have different interests and preferences within the broad area of that specific conference. Therefore depending on the scope of such an event, a mobile multimedia recommender system is necessary to generate efficient, trustworthy and reliable sessions of the conference for attendees/participants.

*C. Context-Aware Services*

In contrast to non-mobile situations such as desktop computer use, mobile computing has brought the notion that the physical and logical context of users or participants from the smart community can power the performance of services they demand for. A fascinating aspect of computing with mobile devices is the decomposition of task from the location and situation where it is performed in the community [29][30].

Users can network with coworkers in real-time from home,

in meetings, conferences and tradeshows, at the airport, in the office, in the supermarket etc. In broader terms, context may refer to any aspect of circumstances where an entity (person, place, or computing object) may invoke computing functions. With current mobile applications, functions would most likely to be specified as demands for service from either local or distant service providers. In this kind of condition, the mobile user and his/her device becomes the service consumer.

The digital representation of contextual data can avoid the need to track down the information in other ways. For instance, a location-aware museum guide is more effective than a pamphlet one must search through. An automatically recorded meeting eliminates the need to take exhaustive notes or ask someone who attended. Such familiar situations point to the great interrelatedness of people, events and the compactness of the contexts in which we function.

As depicted in Figure 3, any given context may comprise information about the physical world (e.g., location, movement, device characteristics, etc.) and about the logical domain neighboring the service consumer. The logical world contains information about personality, preferences and relationships in different domains, such as friends, work, family and others. Even historic information about any of these features might be incorporated. For example, home address might be contained in the service consumer's logical context. Context data about the physical world can be collected in real-time from sensors available in a mobile device.

But the necessary information about the logical world is collected directly from the user/participant or gathered from interactions and communications the user has made with service providers [31]. Whatever the nature of such information, context may come from different ways and has a relatively temporary lifetime. Context-aware smart systems have been studied in various services, for instance, event navigation [32], tourist guidance [33][34], and context-aware messaging [35]-[37].

### III. TAXONOMY OF MOBILE MULTIMEDIA RECOMMENDER SYSTEMS

#### A. General Architecture

Generally, recommender systems suggest items of interest to users based on, for example, their explicit and implicit preferences and the preferences and interests of other users through user and item attributes. Recommender systems share in common a means for describing the items that may be recommended, a means for creating a profile of the user that describes the types of items the user likes, and a means of comparing items to the user profile to determine what to recommend. The user profile is often created and updated automatically in response to feedback on the desirability of items that have been presented to the user.

A mobile device user can acquire relevant multimedia interests and preferences through a mobile multimedia recommender system based on the same scenarios described above. Mobile multimedia recommender systems usually perform three main functions, including: information collection

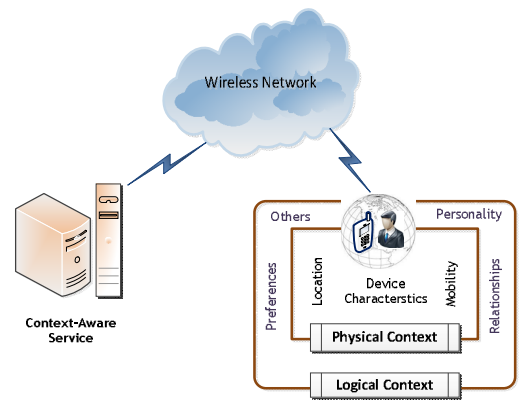

Figure 3. A context-aware system

(explicit or implicit feedback), recommendation learning process (learning algorithm and information filtering) and resource prediction/recommendation. The remaining part of this section explains these functions one by one.

*1) Information Collection*

Mobile multimedia recommender systems through explicit and implicit feedback methods collect all the user interests pertaining to multimedia information for the prediction task including the users' attributes, behaviors, the content of the resources the mobile user accesses as well as context such as location, time or weather.

*Explicit feedback*: In explicit feedback, the user of the recommender system provides data willingly. Generally, users are required to fill mobile device interface forms at the beginning of a sign up process of the system, so that the system can learn a user's model and create a user profile. These forms, depending on the recommender system, usually require the filling of basic demographic information such as age, gender, education, occupation, location or user interests. A mobile academic video recommender system will for example require the location of the learner, educational interest of the learner, as well as his/her learning styles and cognitive abilities towards learning. In a movie recommender system such as *MovieLens*, the user can state interests as "I like comedy films" or "I don't like action films"

*Implicit feedback*: In implicit feedback, the user is not aware of the fact that he/she is providing feedback and his/her behavior is being observed by the mobile multimedia recommender system. Implicit feedback can be gathered by monitoring the user activity and behavior. For instance, in a conference, a mobile multimedia recommender system can keep the list of conference sessions that the user participated in or even better, it can be thought that if a user A, participates in more sessions of a similar conference event CE, then a prediction can be made that "A is interested in CE". In implicit feedback, users are not engaged, but the gathered results might not be as relevant as the results that are collected from explicit feedback. Another example could be if a user watches the whole duration of a mobile academic video that is a sign that the user likes such mobile academic videos and vice versa.

In general, there are tradeoffs between implicit and explicit feedbacks, since implicit feedback methods can collect a large amount of data with some uncertainty as to whether the user actually likes the item. Explicit and implicit data acquisition methods present advantages and disadvantages: explicitly acquired data are more accurate in expressing preferences or interests while implicit methods are considered unassuming to user's main goal in using a mobile multimedia recommender system.

*2) Learning Process*

After (multimedia) information collection, a mobile multimedia recommender system then applies its learning algorithm to filter and exploit the users' features from the collected information. Algorithms for the learning process, which are elaborated more in subsequent section below, differ. Collaborative recommendations can be grouped into two general classes: memory-based and model-based. Memory-based (or correlation-based) techniques apply the user database to calculate the similarity between users or items and make predictions or recommendations according to those calculated similarity values. Model-based techniques apply the user database to estimate or learn a model to make prediction. The model can be a data mining and machine learning algorithm, such as Bayesian belief nets, association rules or neural networks. Context-aware algorithms also filter and exploit the features of the user through available context such as location and time.

*3) Resource Prediction*

After the learning process, predictions on what kind of multimedia resources the user may prefer are then made directly based on the dataset collected in the information collection and learning process stages. A prediction could be based on item similarity, user similarity or context with a reflection of user preferences and interests. Figure 4 depicts the framework of a mobile multimedia recommender system in which information about user's multimedia interests are collected from the user through explicit and implicit procedures, after which the system uses a relevant recommendation algorithm to filter personalized and specific multimedia for the user and finally makes a multimedia prediction to be recommended for the user.

*B. Classification*

*1) Collaborative Filtering (CF) Recommender Systems*

In this category, users are recommended items that people with similar tastes, interests and preferences liked in the past. Through collaborative filtering algorithms, collaborative recommender systems identify users who share the same preferences (e.g., rating patterns) with the active user, and propose items (i.e., multimedia) which the like-minded users favored (and the active user has not yet seen).

Collaborative or social recommender systems, which are the most well-known type of recommender systems do not have the overspecialization drawbacks which are dominating content-based recommendations, but have their own limitations such as cold-start (new user), first rater (new item), rating sparsity, scalability, reduced coverage, synonymy, neighbor transitivity, black sheep, gray sheep and shilling attacks [38]. Generally, CF uses two main approaches, namely, user-based

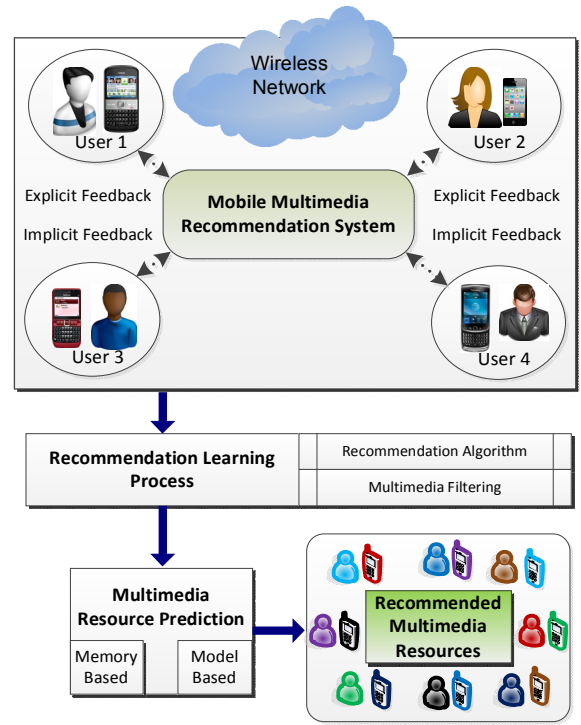

Figure 4. Mobile multimedia recommendation system framework

CF and item-based CF.

*2) Content-Based Recommender Systems*

In this category, recommendations are provided by automatically matching a user's interests with item contents. Items that are similar to those the user preferred in the past are effectively recommended. Note that recommendations are made without relying on information provided by other users, but solely on items' contents and users' profiles. The more descriptive they are the more accurate the prediction is. In content-based filtering only very similar items to previous items consumed by the user are recommended which creates a problem of overspecialization since there may be other items which are relevant and can be recommended but because they haven't been rated by the user before, recommendation becomes impossible [39].

*3) Hybrid Recommender Systems*

Both recommender systems described above exhibit challenges, advantages and disadvantages. Significant research effort has been devoted to hybrid recommendation methods that combine collaborative and content-based filtering as well as other recommendation methods. Burke [40] outlined the different hybridization strategies and methods for hybrid recommender systems. This is elaborated in Table I.

*4) Knowledge-Based Recommender Systems*

The knowledge-based approach is another type of recommendation system. This approach aggregates the knowledge about the users and items (multimedia), and then applies this knowledge to generate recommendations. Knowledge-based recommender systems do not attempt to build long-term generalizations about their users, but rather based on demographic classes such as gender and age. For example, the evaluation procedure used in [10], utilized a demographic recommendation algorithm as part of the integration of three other recommendation algorithms to provide a hybrid recommendation as a prototype.

TABLE I
HYBRIDIZATION METHODS

| Hybridization Method | Description |
|---|---|
| Switching | The system switches between recommendation techniques depending on the current situation. |
| Mixed | Recommendations from several different recommenders are presented at the same time. |
| Weighted | The scores (or votes) of several recommendation techniques are combined together to produce a single recommendation. |
| Cascade | One recommender refines the recommendations given by another. |
| Feature combination | Features from different recommendation data sources are thrown together into a single recommendation algorithm. |
| Meta-level | The model learned by one recommender is used as input to another. |
| Feature augmentation | Output from one technique is used as an input feature to another. |

TABLE II
COMPARISON OF TRADITIONAL RECOMMENDER SYSTEMS

| Recommender System | Advantages | Disadvantages | Requirements | Technical Aspects |
|---|---|---|---|---|
| Collaborative Filtering [38][40]-[44] | No domain knowledge is required. | First Rater issues, Cold-Start and Sparsely problems. Insensitive to preference change. | Needs a set of users with interests and preferences. Needs large historical data set. | Easy to create and use. Makes Recommendations based on the past interests of the user. |
| Content-Based Filtering [39][40][46] | No domain knowledge is required. | Overspecialization. | Needs knowledge about user's preferences. | Considers the preferences of a single user to make recommendations. |
| Knowledge-Based [40][46]-[48] | Sensitive to preference change. Does not need to be initialized with a database of user preferences. | Knowledge acquisition. | A very good understanding of the domain in question. | Knowledge engineering expertise. |
| Utility-Based [45] | Can feature non-product attributes, such as vendor reliability and product availability. | Creating a utility function for each user. | Needs to know the utility functions of the user in order to generate a user profile | Makes suggestions based on a computation of the utility of each object for the user. |
| Demographic [10][40] | Not likely to require a history of user ratings as required by collaborative and content-based techniques. | Demographic information in a user model can vary greatly, because of different types of systems which require recommendations. | Needs to know personal attributes and demographic classes of users. | Aim to categorize the user based on personal attributes and make recommendations based on demographic classes such as gender and age. |

Table II shows a comparison of traditional recommender systems. Figure 5 depicts a typical architecture of a mobile multimedia recommender system involving collaborative filtering, content-based filtering and hybrid recommender systems. The user model and user's mobile device are all connected to a server through an interactive cloud network. To generate Content-Based Recommendations (CBR) the user's model which comprises his/her rating history and profile is matched with the multimedia information stored in the server through a Content-Based Filtering (CBF) algorithm. To generate Collaborative Recommendations (CR) the user is recommended multimedia information from the server based they prefer to generate a recommendation based on matching between user's need, preferences and set of items available. Knowledge-based recommender systems can answer to the question of how special items can meet the special user's need by applying knowledge [46]-[48].

*5) Utility-Based Recommender Systems*

Utility-based recommender systems make suggestions based on a computation of the utility of each object for the user. The central problem is how to create a utility function for each user. The user profile therefore is the utility function that the system has derived for the user, and the system employs constraint satisfaction techniques to locate the best match.

*6) Demographic Recommender Systems*

Demographic recommender systems aim to categorize the user based on personal attributes and make recommendations on matching between a user's rating history and other user's rating history with similar interests through a CF algorithm. In the Recommendation Agent CBR is combined with CR to generate a Hybrid Recommendation (HR) and thus final multimedia recommendations. The user's Personal Content Manager (PCM) in his/her mobile device manages his/her final multimedia recommendations in the Client Agent which is connected to the Recommendation Agent.

Besides these categories, one special and upcoming type of recommender systems worth noting is Context-Aware Recommender Systems (CARS). Since the general concept of context is very broad, Adomavicius and Tuzhilin [49] noted that when applying CARS there should be a focus on fields such as data mining, e-commerce personalization, databases, information retrieval, ubiquitous and mobile context-aware systems, marketing and management which are directly related to recommender systems.

Traditionally, recommender systems deal with applications having only two types of entities, users and items, and do not put them into contextual information, such as time, location or the company of other people. Examples of applicable contextual information for recommendation, include context-aware recommender systems that suggest queries for a given partial query [50], contextualized media delivery systems [14][51][52] and intelligent tutorial guides [53].

In many applications, such as recommending a trip package, personalized content on a web site, or a movie/TV/video, it may however not be sufficient to consider only users and items – it is also important to incorporate the contextual information into the recommendation process in order to recommend items to users in circumstances such as ubiquity and mobility [49][53]-[55].

*C. Core Algorithms*

Mobile multimedia recommendation algorithms are employed to learn the interests, preferences, tastes and context of mobile device users through matching procedures. Notable core algorithms used in mobile multimedia recommendation include, memory-based CF algorithms, model-based CF algorithms, contextual modeling algorithms and relevance feedback and Rocchio's algorithm [38][39][56].

*1) Memory-Based CF Algorithms*

Memory-based algorithms are typically for CF. They are heuristics that make rating predictions based on the entire collection of items previously rated by users. Memory-based algorithms are also called direct neighbor selection algorithms and have a function of keeping users linked in the network (with a non-null similarity value). A prevalent memory-based algorithm uses the following steps:

- Calculates the similarity or weight $W_{ij}$, which reflects distance, correlation, or weight between two users or items $i$ and $j$.
- Produces a prediction for the active user by taking the weighted average of all the ratings of the user or item on a certain user or item, or using a simple weighted average.

Similarity computation between items and users is a critical step in memory-based collaborative filtering algorithms. Notable types of similarity computations used in memory-based collaborative filtering algorithms are: Pearson Correlation and Vector Cosine-Based Similarity.

- *Pearson Correlation-Based CF Algorithm*: The Pearson measures the extent to which two variables i.e. users or items linearly relate or are similar to each other.
- *Vector Cosine-Based Similarity*: The similarity between two documents can be measured by treating each document as a vector of word frequencies. This approach can be adopted in CF by using users or items instead of document ratings of word frequencies.

*2) Model-Based CF Algorithm*

Model-based algorithms use the collection of ratings to learn and compute a model of a user, which is then used to make rating predictions. If a comparison is made between memory-based algorithms and model-based algorithms, memory-based algorithms can be considered "lazy learning" algorithms and methods because they do not build a model, but instead perform the heuristic computations when collaborative recommendations are requested. The design and development of models such as data mining and machine learning algorithms can allow the system to recognize complex patterns based on training data, and make intelligent predictions for the CF tasks for real-world data or test data based on the learned models.

Model-based CF algorithms such as Bayesian Models, Clustering Models and Dependency Networks have been investigated to address the disadvantages of memory-based CF algorithms. In both memory-based and model-based approaches used in CF, the users are clustered based on the similarity. Similarity is calculated based on rating matrix, using Pearson correlation function or Cosine function.

- *Bayesian Classifier/Simple Bayesian CF Algorithm*: The Bayesian classifier which is commonly used and applied in CF systems has been proved as an efficient classification mechanism in several fields. In principle the classifier bases on a 2-class decision model, where an item in a domain object belongs to the class like or dislike. Some researchers have recognized Bayesian classifier as an exceptionally well-performing model-based CF algorithm. For instance, Pessemier *et al.* [57] reported that the structure of the user profile, the user interaction mechanism, the recommendation 8algorithm and an improved version of the Bayesian classifier that incorporates aspects of the consumption context (like time, location, and mood of the user) can make the suggestions and predictions more accurate. To this effect, they developed a methodology for mobile devices that makes the huge content sources more manageable by creating a user profile and personalizing the available resources.

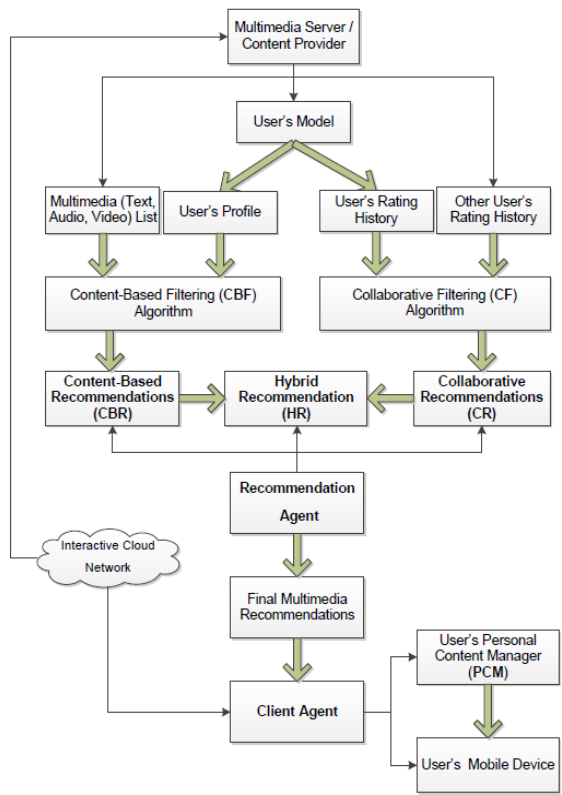

Figure 5. Workflow of hybrid mobile multimedia recommendation

- *Bayesian Belief Network CF Algorithm*: A Bayesian Network (BN) is a Directed Acyclic Graph (DAG), where the nodes represent the variables from the problem that has to be solved. In the application of BN, the first task is to select those variables which are relevant to the problem being tackled. Each variable will be a node in the DAG and whenever two variables are related, a path must exist between them in the graph. These connections can be determined from an input dataset by means of a learning algorithm. The Bayesian network is commonly used for CF [58].

*3) Vector Space Model*

The vector space model is used to generate content-based recommendations. The vector space model can also be used in collaborative filtering to gather documents on basis of interest of community as a whole. The basic concept of the vector space model is the selection of an item vector that is most similar to a retrieval key. In such scenarios, there is the need to select one retrieval key and several item vectors.

For example, in a mobile social learning video recommender system, the item vectors are the number of academic videos and the retrieval key is the mobile social learner's interest and preference. There exist several methods such as the inner product and the least-square method that can be used to calculate the similarity of users and items.

*4) Relevance Feedback and Rocchio's Algorithm*

Methods that help users to incrementally refine queries based on previous search results have been the focus of much research. These methods are commonly referred to as relevance feedback. The Rocchio's algorithm, a widely used relevance feedback algorithm that operates in the vector space model, is based on the modification of an initial query through differently weighted prototypes of relevant and non-relevant documents. The approach forms two document prototypes by taking the vector sum over all relevant and non-relevant documents.

*5) Contextual Modeling, Pre-Filtering and Post-Filtering Algorithms*

The contextual modeling approach of mobile multimedia recommender systems uses contextual information directly in the recommendation function as an explicit predictor of a user's rating for an item. The contextual modeling approach gives rise to truly multidimensional recommendation functions, which principally represent predictive models that are built using probabilistic models, regression, decision trees or other technique or heuristic calculations that incorporate contextual information in addition to the user and item data, i.e., *Rating = R*(*User*, *Item*, *Context*). The contextual pre-filtering and post-filtering approaches can however use traditional two dimensional (2D) recommendation functions.

The three paradigms for context-aware recommender systems (contextual modeling, pre-filtering and post-filtering) offer several different opportunities for employing combined approaches just like combination of other traditional recommender systems. One possibility is to develop and combine several models of the same type.

For example, Adomavicius *et al*. [55] followed this approach by developing a technique that combines information from several different contextual pre-filters. Table III summarizes

TABLE III
SOME ALGORITHMS THAT CAN BE USED IN MOBILE MULTIMEDIA RECOMMENDER SYSTEMS

| Recommendation Algorithm | Recommendation Algorithm Techniques | Recommender System | Background/Entities/Input | Process |
|---|---|---|---|---|
| Memory-Based Collaborative Filtering [38][56] | Pearson Correlation-Based and Vector Cosine-Based Similarity | Collaborative Filtering | users and items | Makes ratings predictions based on the entire collection of items previously rated by users through a similarity function. |
| Model-Based Collaborative Filtering [38][56] | Bayesian Classifier/Simple Bayesian/Naïve Bays , Naïve Bayes Extended Logic Regression (NB-ELR), Tree Augmented Naïve Bayes Extended Logic Regression (TAN-ELR) and Bayesian Belief Network (BAN) | Collaborative Filtering | users and items | Uses the collection of ratings to learn and compute a model of a user through a similarity function. Predicts through probability methods, the level of interest of a particular user in a particular domain using conditional probabilities. |
| Content-Based [39] | Vector Space Model | Content-Based Filtering and Collaborative Filtering | users and items | Selects an item vector that is most similar to a retrieval key when there is the need to select one retrieval key and several item vectors. |
| | Relevance Feedback and Rocchio's Algorithm | Content-Based Filtering | users and items | Modifies an initial query of a user through differently weighted prototypes of relevant and non-relevant documents. |
| Context-Aware [49][53][55] | Contextual Modeling, Pre-Filtering and Post-Filtering Algorithms | Content-Aware | users, items and context | Uses item and contextual information in the recommendation function to make a prediction of a user's rating for an item. |

some of the existing algorithms that can be used for mobile multimedia recommendation.

*D. Performance Metrics*

Several metrics have been used for evaluating recommender systems and can broadly be categorized into predictive accuracy metrics, classification accuracy metrics, and rank accuracy metrics [59][60].

*1) Predictive Accuracy Metrics*

The predictive accuracy metrics measure how close the recommender system's predicted ratings are to the true user ratings. Predictive accuracy metrics are particularly important for evaluating tasks in which the predicted rating will be displayed to the user, e.g. annotation in context. For example, the *MovieLens* movie recommender predicts the number of stars that a user will give each movie and displays that prediction to the user. Predictive Accuracy Metrics will evaluate how close *MovieLens*' predictions are to the user's true number of stars given to each movie. Even if a recommender system was able to correctly rank a user's movie recommendations, the system could fail if the predicted ratings it displays to the user are incorrect.

*2) Classification Accuracy Metrics*

The classification accuracy metrics measure the frequency by which a recommender system makes correct or incorrect decisions about whether an item is good. Classification metrics are thus appropriate for tasks such as "find good items" when users have true binary preferences. Several recommender systems were developed with a more direct focus on actual recommendation.

*3) Rank Accuracy Metrics*

The rank accuracy metrics measure the proximity between the ordering predicted by a recommender system to the ordering given by the actual user, for the same set of items available. They measure the ability of a recommendation algorithm to produce a recommended ordering of items that matches how the user would have ordered the same items. Unlike classification metrics, ranking metrics are more appropriate to evaluate algorithms that will be used to present ranked recommendation lists to the user, in domains where the user's preferences in recommendations are non-binary.

*E. Challenges*

A number of challenges remain for mobile multimedia recommendation. Below we will discuss some of them in the context of smart communities.

*1) Mobile Device Limitations and Connectivity*

Mobile device challenges and limitations such as limited display/screen sizes, low battery lives, limited processing power, limited storage and limited input capabilities affect their usability practices. Recent advances in microprocessors architectures and high bandwidth networks permit one to consider high performance Peer-to-Peer (P2P) distributed computing as an economic and attractive solution to mobile multimedia recommender systems [61]-[63]. For example, Chen *et al.* [64] designed and optimized Bloom Filter settings in a peer-to-peer mutli-keyword searching which requires intersection operations across Wide Area Networks (WANs). Another technical challenge in this regard is sustainable network connectivity. The mobility of such recommender systems requires efficient, effective and reliable network technologies for sustainability.

*2) Mobile Context Acquisition*

Fast and accurate acquisition of user's preferences as well as context in mobile multimedia recommender systems is still a challenge. The acquisition of contextual information in mobile multimedia recommender systems is mainly done through a manual explicit procedure. Requesting such manual inputs from mobile device users may hamper mobile multimedia recommendations if a user is not willing to provide relevant and needed contextual information or provides false contextual information.

*3) Mobile Context Depiction*

Due to the lack of a standard representation for contextual data in mobile devices, sharing and reuse of data among users is a challenge. The representations of contextual information in mobile devices are diverse which make contextual collaboration difficult and hence affects mobile multimedia recommendation.

*4) Dataset Acquisition and Sharing*

The acquisition and sharing of relevant datasets for mobile multimedia recommender systems for testing of algorithms in scientific experiments remain a challenge. In many mobile computing disciplines a large collection of multimedia datasets are important for effective evaluation of novel algorithms. Some standard mobile multimedia datasets for recommendation are private (cannot be used commercially) and require permission from the owner before they can be used. Such permissions are accompanied with dataset restrictions and sometimes attract expensive costs. Sharing of mobile multimedia datasets among researchers is also a challenge due to issues of privacy of the owner and the owner not wanting to release the dataset for public use.

*5) Evaluation*

In order to prove successful application of algorithms involving users, items and context in mobile multimedia recommender systems, the evaluation technology need to be strengthened. For example, the promotion of evaluation studies and strategies that assess the impact of individual context elements such as location, activity and time in a mobile multimedia recommendation process will improve and strengthen evaluation procedures.

*6) Interaction*

The development of user interfaces for enhancement of interaction between the users and the mobile multimedia recommender system needs to be addressed. It is also important that a user understands why a particular multimedia recommendation is generated through the mobile device for him/her i.e. the reason of a generated mobile multimedia recommendation should be explained to the user for interaction enhancement.

## IV. STATE-OF-THE-ART MOBILE MULTIMEDIA RECOMMENDER SYSTEMS FOR SMART COMMUNITIES

Existing recommendation techniques have been quite successful in commercial domains. Examples of such applications include recommending movies by *MovieLens* and books, CDs, and other consumer products at *Amazon*. Due to the relevance, importance and advantages of mobile device usage in education, events and context-aware situations, as we mentioned previously, our focus in this section is to review existing research and related work in these research areas and also discuss recommendation systems in smart communities comprising of mobile social learning, mobile event guides and context-aware services.

*A. Recommendation in Mobile Social Learning*

Recommender systems in the mobile social learning and Technology Enhanced Learning (TEL) have been investigated extensively by researchers during the last decade [53][65]-[71]. When suitable resources from a potentially overwhelming variety of choices are identified, recommender systems offer an encouraging approach to facilitate both learning and teaching tasks. As learning is taking place in exceptionally diverse and rich environments, the incorporation of contextual information about the learner in the recommendation process has attracted major interests among researchers. Such contextualization is currently being researched as a paradigm for building intelligent systems that can better predict the needs of users (learners), and act more efficiently in response to their behavior.

It is difficult to express a specific learning requirement through keywords. Search engines such as Google do a reduced job when a learner needs content about "relativity theory", oriented to high school level, with a duration of about 40 minutes. Finding relevant resources can be even more difficult when requirements are not always fully known by the learner, such as his/her level of competence or adequate technical format. Recommender systems for learning try to address these challenges - i.e. they attempt to filter content for different learning settings. Apart from using learner interest and available multimedia learning resources as a basis to generate recommendations in mobile social learning, the notion of context has started to attract significant research attention. A learning context describes the current situation of a learner related to a learning activity, and continually changes in a mobile learning environment. As a result, context-awareness in mobile learning has become an essential part when designing more adaptive mobile learning systems.

A new set of recommender systems for learning have been developed in recent years to demonstrate the potential of contextual recommendation. A first example of a context-aware recommender system for learning considers the physical conditions, location of the user and the noise level at this location as a basis to suggest and recommend learning resources to the learner.

Beham *et al.* [70] discussed and addressed the requirements for user models for a Work Integrated Learning (WIL) situation. They presented the APOSDLE people recommender service which is based on an underlying domain model, and on the APOSDLE user model. The APOSDLE People Recommender Service works on the basis of the Intuitive Domain Model of expert finding systems, and explain to support interpersonal help seeking at workplaces. Broisin *et al.* [71] used location and tracking as context and presented a solution for recommending documents to students according to their current activity that is tracked in terms of semantic annotations associated to the accessed resources. Their approach was based on an existing tracking system that captures the user's current activity, which is extended to build a user profile that comprises his/her interests in term of ontological concepts. A recommendation service was elaborated by implementing an algorithm that is alimented by Contextualized Attention Metadata (CAM) comprising the annotation of documents accessed by learners.

In the same vein, Yin *et al.* [72] proposed a system to address and contribute to the challenge of location in a mobile social learning environment based on hypothesis that involves asking for help from others. These hypotheses were: 1) the closer people are, the easier it is to get help; 2) the more simple things are, the more it is easier to get help with them. They carried out a survey to examine the above hypotheses. The results showed that these hypotheses are correct. Based on these hypotheses, they proposed a social networking service based on a mobile environment called *SENSMILE*, which supports learners to find a partner who can solve their problems at the online community, and an appropriate request chain of friends will be recommended upon their request by utilizing personal relationships. The system also supports collaborative learning by using location based sensing information.

Chen *et al.* [73] developed a Learning Companion Recommendation System (*LCRS*) on Facebook. *LCRS* supports a mobile collaborative learning by automatically collecting profile data such as interests and professional abilities of friends', in accordance to their learning needs. Furthermore, the technology acceptance model and partial least squares regression are used to investigate learners' acceptance of the LCRS for learning activities.

To implement the client-side recommendation model, Kim *et al.* [74] proposed "buying-net", a customer network in ubiquitous shopping spaces. Buying-net is operated in a community, called the buying-net space, of devices, customers, and services that cooperate together to achieve common goals. The customers connect to the buying-net space using their own devices that contain software performing tasks of learning the customers' preferences, searching for similar customers for network formation, and generating recommendation lists of items. Buying-net attempts to improve recommendation accuracy with less computational time by focusing on local relationship of customers and newly obtained information.

Zhu *et al.* [75] presented a user-centric system, called *iScope*, for personal image management sharing on mobile devices. The *iScope* system uses multimodality clustering of both content and context information for efficient image management and search, and online techniques for predicting images of interest. *iScope* also supports distributed content-based search among networked devices while maintaining the same intuitive interface, enabling efficient information sharing among people.

Cui and Bull [76] discussed how to support the mobile language learner using a handheld computer. They introduced *TenseITS*, a language learning environment that adapts the interaction to the individual learner's understanding, as represented in a learner model constructed during the interaction. *TenseITS* also adapts according to contextual features of the learner's location that may affect their ability to study.

Yau and Joy [77] also introduced a context-aware and adaptive learning schedule framework which makes use of a learning schedule to support the students' daily routines, adapts the activities to the student's learning styles and then selects the appropriate activity for the learner based on their current learning context.

*B. Recommendation in Mobile Event Guide*

Social activities are occurring all over the world as an important part of life and an activity imminently requires answering the critical questions such as "what", "how", "where", "when", "why" and "who". Even if it is hard to give information and recommend events that may interest people regardless of the above questions, systems and methods have been developed for selecting and recommending events to a user of media content based on information known about the user in the community, including the user's interests, consumption history and preferences.

An intelligent recommendation system can be used in a social networking site in order to recommend people according to content and collaboration assessment. The system further includes an event database containing information about upcoming events. When a user interacts with the system, such as to access a media content item, the user is notified of one or more events based on the user's history. In this way users are automatically provided with updated event recommendations based on the known interests of the user, without the need for the user to subscribe to an event recommendation service and maintain the subscription to accurately reflect the user's current interests.

By the time one can expect user feedback on a specific event, that event is no longer relevant: an event recommendation system therefore has to recommend items for which no explicit feedback exists. This distinguishes events from other information items, like movies, for which some user feedback is directly available and continues to be useful. In general, one can approach event recommendation using existing content-based methods, relating event descriptions to user preferences. However, the quality of content-based recommendation is often highly dependent on the amount of user feedback available. We expect feedback about events to be relatively scarce; events are often topically diverse and new events may have low similarity to prior events.

Based on these, Kayaalp *et al.* [78] recommended events to users within a social networking site. It can be any networking environment. They used a social environment that has been designed similarly to Facebook. Their application has also been integrated with several social networking sites, like Twitter or LinkedIn. Their system also permits users rating events they have attended, organized or planned. Given the social network between people, the system tries to recommend upcoming events to users. For this purpose, they used a combination of content-based and collaborative filtering. The work in [79] considered a recommendation system in which the information items are events. The authors pointed out that event recommendation has various potential applications. Park *et al.* [80] studied a location-tracking system, which records the user's whereabouts within a geographical region or inside a building. Given location coordinates, it is desirable to provide users with personalized location-based services.

The study in [81] exploits collaborative filtering to construct a set of reviewing assignments for conferences in response to a primary input of paper-reviewer 'bids' expressing interest or disinterest of reviewers in specific papers. While previous works apply collaborative filtering in "traditional" settings, based on common pools of items and users, they solved a different problem, where the sets of known items and the items for which recommendation takes place are distinct. In particular, their approach can be seen as an extension of event recommendation in [82] where, in contrast to the more common user-item rating scenario, the authors seek to rank future events on the basis of their word content. As a result, their problem formulation is closely related to the low-rank document retrieval method of [83].

The task of recommending live events, such as TV shows, has been already investigated in the past years. Basso *et al.* [84] showed how to recommend live events to users without any knowledge about the broadcast content and user's interests. Recommendations can be given both globally and personally. It is important to underline that the most popular events are easier to predict since users tend to naturally focus on them, even without any specific suggestion. On the contrary, granting a high novelty in personal recommendations is a more challenging goal due to the reduced amount of explicit information. Using mobile phone data, Quercia *et al.* [85] have been able to infer which events had been attended and have considered the attendance as implicit user feedback. In studying different ways of recommending events, they found that recommending nearby events is ill-suited to effectively recommend places of interest in a city. In contrast, recommending events that are popular among residents of the area is more beneficial. They put forward the idea that, by sharing attendance at social events, people are able to receive quality recommendations of future events.

To avoid using explicit inputs as context sources, Cimino *et al.* [86] proposed and designed an approach based on the emergent paradigm [87] for detecting events. They designed a form of collaborative situation awareness considering an important class of events. In this scenario, they also discussed collaborative multi-agent scheme for the detection of such events, structured into three levels of information processing. The first level is managed by a stigmergic paradigm [88], in which marking agents leave marks in the environment that is associated to users' positioning.

The work in [89] *CityEvents*, a mobile solution that provides real time awareness based information of local events and news. The presented ubiquitous system is a simple and easy to use event guide, providing a location-awareness and content-awareness ordering data to the user. The proposal was evaluated and demonstrated in i-Phone devices and it is ready for use. The study in [90] proposes a mobile ad hoc networking approach for detecting and querying events related to farm animals such as oestrus, animal diseases and decreased efficiency of pastures. This problem is essential because inattention to the health or welfare of the animals can lead to reduced productivity and death of valuable stock. The proposed approach guided by the identified requirements, utilizes the available networking infrastructure but also works in a fully ad hoc infrastructure-less condition.

An interactive learning-guide system, designed with Radio Frequency Identification (RFID) electronic tags and PDA, to assist the visitors in better understanding the exhibitions is proposed in [91]. Visitors obtain the exhibition-related information through a PDA with interactive RFID reading. Unlike traditional systems, this mobile guide is supplemented with data mining, information retrieval and location-awareness to help achieve the ideal format for interactive usage. The system utilizes the user profiles to analyze individuals' viewing records by employing collaborative filtering and Apriori-like algorithms to create recommended viewing rules for visitors.

The study in [92] solved a novel problem of route recommendation which guides the user through a series of locations. The recommendation is made by matching the user's current route with the set of popular route patterns. Sequential pattern mining methods are used to extract the popular route patterns from a large set of historical route records from previous users. As an application with real requirements, the Bar Tour Guide is used to demonstrate the effectiveness of the proposed solution.

Eliciting requirements on-site is challenging as analysts have to simultaneously observe the environment, interact with people and operate requirements engineering tools. Seyff *et al.* [93] explored the use of context-aware technologies to provide better guidance and support for on-site analysts. The context aware mobile scenario presents tool guides to analysts by automatically highlighting scenario events relevant to the currently observed work task. The study in [94] deals with the problem of deriving personalized recommendations for daily sightseeing itineraries for tourists visiting any destination. Their approach considers selected places of interest that a traveler would potentially wish to visit and derives a near-optimal itinerary for each day of visit; the places of potential interest are selected based on stated or implied user preferences. Therein, the authors proposed a heuristic solution to this problem and discussed its implementation.

*C. Recommendation in Context-Aware Services*

The evolution of mobile devices coupled with the ubiquitous availability of wireless communications has promoted the development and commercialization of new and sophisticated mobile services. An example of such mobile services is location-based information services suited for the needs and constraints of mobile users. Because of the development of these technologies and the incredible appeal of mobile devices and services, there has also been much research and development effort in trying to apply recommendation technologies to this field [49][54].

However, recommendation systems and techniques, which have proved to be successful for standard PC users, cannot be forthrightly applied for mobile users. On one hand, mobile multimedia recommender systems have to overcome the impediments typically present in mobile usage environments, i.e., the resource limitations of mobile devices, the limitations of wireless networks, the impacts from the external environment, and the behavioral characteristics of mobile users. On the other hand, mobile multimedia recommender systems have the ability to exploit two peculiar context-aware characteristics of mobile information services:

*1) Location-Awareness*

The first exclusive property is "location-awareness" through Location Based Services (LBS). Location-awareness through LBS is the procedure of discovering and knowing the user's physical position at a particular time. Such discovery and knowledge can be exploited as an important source of information to adapt the information delivered by the system and hence influence the contextual generation of a recommendation.

*2) Ubiquity*

The second exclusive property is "ubiquity". Ubiquity is the ability of mobile multimedia recommender systems to deliver the multimedia information and services to mobile device users wherever they are and whenever they need. Original recommendation classifications and computation models that are applicable in multimedia have been explicitly developed for mobile recommender systems.

In the field of mobile multimedia recommender systems, various types of context such as computing devices and networks, location, physical conditions, temperature, weather, and time are very important and need to be considered and incorporated in a recommendation process. Here we elaborate on some related works and existing research conducted in the area of mobile multimedia recommender systems that focus on context-aware scenarios.

Yong *et al.* [1] proposed a personalized recommendation scheme which considers the activities of the user at runtime and the information on the environment around the user. It allows efficient operation in mobile devices, and interoperability between the TV multimedia metadata and ontology. The accuracy of the proposed scheme was evaluated by an experiment, which revealed a significant improvement compared to the existing schemes.

The study in [95] describes the project *SNOPS*, a smart-city environment based on Future Internet technologies. The authors therein focused on the context-aware recommendation services provided in the platform, which accommodates location dependent multimedia information with user's needs in a mobile environment related to an outdoor scenario within the cultural heritage domain. In particular, they described a recommendation strategy for planning browsing activities, exploiting object features, users' behaviors and context information gathered by opposite sensor networks. Preliminary experimental results, related to user's satisfaction, have been carried out and discussed.

In [96] the comparison of four major recommender system approaches: content-based, content-based with context, collaborative with context and collaborative were studied and used to provide personalized recommendations to museum visitors. Based on the result of the comparisons, the authors implemented a recommender system for a mobile museum guide called *MyMuseum*. Discussions were presented on *My Museum's* functionalities and comparison was made with *MyMuseum* and other systems. The authors also presented results of their testing and concluded that *MyMuseum* is a useful application that museum visitors can use during their tours.

The study in [97] introduced an approach for context-aware personalization of mobile multimedia services. The authors developed a generic framework for application developers that can easily be configured and extended with application-specific algorithms matching content and context. In the proposed system, content selection is performed in a distributed way. The study in [98] presented in detail a platform implementation for enabling the creation of lightweight content- and context-aware mobile multimedia services. The platform supports content management functionalities in order to enable easy creation of specialized multimedia services for various target groups and purposes. The solution includes context metadata support for mobile multimedia content and creation of location-aware multimedia services.

In [99] a new framework for context-aware recommendation of visual documents by modeling the user needs, the context, and the visual document collection together in a unified model was proposed. The user's needs for diversified recommendations were addressed. The pilot study showed the merits of the approach in using content based image retrieval. To provide media recommendations for smart phones based on all three context categories, the authors of [100] presented a generic and flexible $N$times$M$-dimensional (N2M) recommendation model. The model considers context information ranging from user preference and situation to device and network capability as input for both content and presentation recommendations.

## V. DISCUSSION AND OPEN ISSUES

In the previous section of this paper some existing research of mobile multimedia recommendation in three smart communities were surveyed. The survey delved into contextual scenarios and other relevant content-based and collaborative filtering recommendations. Each mobile multimedia recommendation system had its own strength and weakness. The survey involved in the recommendation in mobile social learning revealed that CARS had been applied to most of the existing research and in some cases context had been combined with collaborative filtering and/or content-based filtering. Location, physical conditions as well as time were the types of contexts that were prevalent among the various research surveyed. These applications can however be improved in terms of standard context acquisition or sensing. Incorporating context such as emotion and computing (mobile) is essential, so that the current mood (good or bad) of mobile social learners is verified before generating a recommendation.

Mobile advertising during event (conference, tradeshows, seminars etc.) guides is very important for participants to express themselves further and improve event collaboration. Our survey reveals that most of the existing research does not incorporate mobile advertising between event participants. To facilitate the discussion of this paper, we came out with the following subsections consisting of relevant open issues based on the survey conducted.

### A. Predictive Models for Evaluation in Mobile Multimedia Recommendation

After the identification of significant sets of contextual conditions, a predictive model, such as kernel-based classification that can predict how the evaluation of mobile multimedia changes as a function of the contextual factors is a necessity for research and development in mobile multimedia recommendation. The predictive model can be used to select mobile multimedia given a target context. The step involved in a relevant predictive model in mobile multimedia recommendation requires the collection and utilization of explicit ratings for mobile multimedia under several distinct contextual conditions.

### B. Mobile Multimedia Advertising in Smart Communities

In smart communities such as academic conference events, participants will need/want to advertise and share their personal multimedia resources such as academic papers and presentations with other conference participants who have the same multimedia resource interests and preferences. For instance when a participant of an academic conference or workshop wants to advertise and share his/her academic paper with other attendees who are likely to be interested, a standard mobile multimedia recommender system should be able to generate/suggest a recommendation to the other participants who are likely to be interested in the active participant's academic paper through a mobile advertisement. Such an innovation will improve the sharing capabilities of mobile multimedia resources and enhance social awareness in a smart community.

Assuming there is a social intelligence and technology smart meeting which involves different sessions, how smart meeting participants can be advised as to which session(s) to attend should also be possible by advertising through a mobile multimedia recommender system using contextual factors such as schedule, location and time.

### C. Friendship, Trust and Social-Awareness in Smart Communities

Socializing and familiarization at events such as smart meetings is very important. Through socialization and familiarization, participants of the conference or a workshop become friends and are likely to gain trust for each other in terms of academic knowledge and contribution. A particular participant may be very popular in a specific field and as a result of his popularity. A mobile multimedia recommender should be able to make it possible for participants to be socially aware of each other and make trustworthy friendships. There is therefore the important need of incorporating positive social properties such as social ties and centrality and solving negative social properties such as selfishness, so that more accurate, effective and novel mobile multimedia recommendation can be generated in smart communities. Furthermore, an important issue in mobile multimedia recommendation through friendship is how mobile device users who are friends can gain effective trust for each other. Usually mobile device users who are friends can trust each other, but sometimes that is always not the case, therefore the system should be able to detect such scenarios.

*D. Mobile Multimedia Context Sensing and Depiction*

Several systems described above indicate that they used contextual data for recommendations but do not actually describe into detail how the contextual information was acquired. The development of mobile context sensors that automate the acquisition of the context dimensions in a standard procedure is a priority for mobile multimedia recommendation. The development of context sensors will also prevent explicit procedures of context acquisition, thereby preventing the involvement of mobile device user manual input and supporting the uptake of contextual information in mobile multimedia recommender systems through automated procedures.

After context is sensed and acquired for mobile multimedia recommendation, the standardization of depicting and mapping contextual data to existing standard specifications, so that data complaint pertaining to mobile multimedia recommendation can be exchanged and reused by mobile device users is also very important.

*E. Innovative Mobile Multimedia Recommendation Paradigms*

*1) Proactive and Sensor-Based Recommendations*

Traditional recommender systems usually require a user to submit query explicitly. A mobile recommender system that proactively pushes just-in-time personalized multimedia contents to mobile device users based on their contextual information as well as interests and preferences will not only solve the problem of information overload but also improve recommendation access in mobile multimedia as well as reduce keystrokes by mobile device users. Most mobile multimedia recommender systems wait for a user's request before delivering any recommendation. Proactive and sensor-based recommendations will support a more exploratory approach to information search [101][102].

In proactive and sensor-based recommendations, the recommender system just detects how the user is accessing and using the system, for example browsing and reading habits. Therefore, minimal input is asked from the user, but none of the reviewed systems is capable of proactively interrupting the user's activity with unsolicited but relevant recommendations.

Some researchers [103]-[107] have however recently developed mobile multimedia recommender systems that generate proactive and sensor-based recommendations to some extent. The new generation of inexpensive and reliable biometric sensors can make this type of recommendations possible and convenient. This can revolutionize the role of mobile multimedia recommender systems from topic oriented multimedia information seeking and decision making tools to multimedia information discovery and entertaining companions.

*2) Intelligent User Interfaces and Improving Hybrid Recommender Techniques*

Extensive and continuous researches in the area of Intelligent User Interface approaches of mobile multimedia recommender systems have to be conducted to improve mobile multimedia recommendation in smart communities. Some hybrid mobile multimedia recommender systems such as [108]-[110] can be further improved with standard contextual incorporation and other relevant and appropriate techniques. In cases where a user doesn't get good recommendations through a mobile multimedia hybrid recommender system, additional algorithms for such systems should be developed to satisfy user's recommendation requests through relevant and appropriate techniques.

*3) Recommendation Explanation and Accuracy*

The issue of recommendation explanation and accuracy has received very little attention but is clearly very important for mobile multimedia recommendation scenario. New approaches for the explanation of the recommendations are needed. A stricter incorporation of the recommender systems with other information services could be an approach, but the core problem of limiting the amount of descriptive information associated to the recommendations is still an open issue [54][111].

*4) User's Memory and Issues of Security and Privacy*

A mobile device user must be open to discover new services and at the same time must be prepared to prevent his/her privacy and security from the potential dangers coming from unknown and hostile programs that may exist in a smart community. Mobile device users are always likely to roam from systems to systems and from networks to networks that vary and are very different. In relation to the privacy issue of mobile multimedia recommender systems, it is expected that there will be considerable growth in the future, i.e., supporting personal memories; helping the user to remember personal facts and tasks, and helping users to make the best usage of multimedia information .

## VI. CONCLUSIONS

As we come to the end of this survey paper, it is undoubtedly clear now, that although there are some progresses in the research of mobile multimedia recommender systems in the smart communities elaborated in this paper, the current trends are not very openly shaped or likely to become much clearer in the very near future. We have elaborated on existing research and state-of-the-art of three smart communities, namely: mobile social learning, mobile event guide and context-aware services and also discussed some new paradigms of mobile multimedia recommendation and verified that some issues are yet to be addressed.

Mobile multimedia recommender systems as elaborated in this paper are very important for smart communities. Factors such as multimedia information overload and user preferences/interests as well as mobile device challenges and limitations have necessitated the development of these systems through relevant algorithms and classifications. It is therefore important to overcome the challenges of mobile multimedia recommender systems in smart communities coupled with researching extensively on the open research issues and new paradigms for mobile multimedia recommendation in order to develop suitable multimedia recommender systems for mobile device users.